# CALCULATION OF THE EQUILIBRIUM COMPOSITION OF METALLIC AND OXIDE MELTS DURING THEIR INTERACTION

## M. Zinigrad

*Ariel University, Israel*

Thermodynamic equilibrium can be sometimes reachedat the interaction between metal and oxide melts in high temperature welding and metallurgical processes. Calculation of equilibrium phase composition is also one of the stages (along with kinetics analysis) of building an end-to-end physical and chemical model of real processes.

Developing a thermodynamic model ofhigh temperature processes is described in many scientific sources (1-18). Thermodynamic patterns of oxidation-reduction reactions on the metal-slag melt interface are also taken into account when building phenomenological models of real welding, facing, and metallurgical processes (19-30).

This paper deals with calculating an equilibrium phase composition for a specific reaction. A method proposed herein also takes into account an interaction among any number of chemical reactions in the metal-oxide melt interface that form an equilibrium phase composition.

Let us first start with a simple case of calculating manganese concentration in steel at 1700ºC, when in equilibrium with the following oxide melt (mas. %):

36.6 - $SiO_2$; 16.4 – $Al_2O_3$; 25.3 – $MnO$; 1.6 – $FeO$; 20.1 – $CaF_2$.

The following chemical reaction to determine manganese contents in steel

$$[Mn] + (FeO) \rightarrow [Fe] + (MnO) \qquad (1)$$

reaches an equilibrium.

This is one of the chemical reactions in the metal-oxide melt interface (Fig. 1). 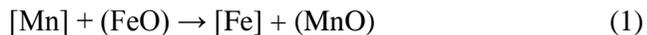

In a first approximation, we will determine manganese equilibrium contents disregarding any other processes. In such a case, our task can be reduced to an equilibrium thermodynamic constant formula for process (1):

$$K_{Mn} = \frac{a_{MnO}^{eq} \cdot a_{Fe}^{eq}}{a_{Mn}^{eq} \cdot a_{FeO}^{eq}}, \quad (2)$$

where $a_{MnO}$, $a_{Fe}$, $a_{Mn}$ and $a_{FeO}$ – are reactant activities in contacting phases.

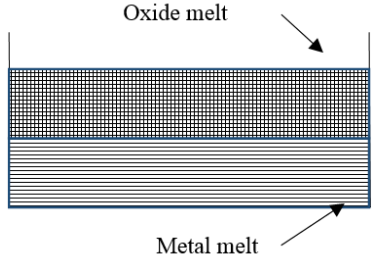

Fig. 1. Reaction scheme.

Let us assume that iron activity equals $a_{Fe} = 1$, as its concentration is close to 100%. By substituting activities with activity factors($\gamma_i$) and concentrations ($x_i$)

$$a_i = (\gamma_i) \cdot (\chi_i),$$

we obtain the following formula:

$$K_p = \frac{\gamma_{MnO} \cdot \chi_{MnO}^{eq}}{\gamma_{FeO} \cdot \chi_{FeO}^{eq} \cdot \gamma_{Mn} \cdot \chi_{Mn}^{eq}} \quad (3)$$

where $\gamma_{MnO}$ and $\chi^{eq}_{MnO}$, $\gamma_{FeO}$ and $\chi^{eq}_{FeO}$, $\gamma_{Mn}$ and $\chi^{eq}_{Mn}$ - are activity factors and molar fractions of reactants in the state of equilibrium.

If we use the following values for our calculations,

$$K_p = 8.3, \quad \frac{\gamma_{FeO}}{\gamma_{MnO}} = 2.5, \quad \gamma = 1.3$$

this equation (3) will look as follows:

$$8.3 = \frac{\chi_{MnO}^{eq}}{\chi_{FeO}^{eq} \chi_{Mn}^{eq} \cdot 2.5 \cdot 1.3}. \quad (4)$$

When taking into account molecular masses ($M_i$), mass % concentrations (% $i$) and the total molar volume of all oxide melt components ($\Sigma n_i$), FeO and MnO molar fractions are determined as:

$$\gamma_{FeO}^{eq} = \frac{n_{FeO}}{\sum n_i} = \frac{\% FeO_{eq}}{M_{FeO} \cdot \sum n_i}$$

and

$$\gamma_{MnO}^{eq} = \frac{n_{MnO}}{\sum n_i} = \frac{\% MnO}{M_{MnO} \cdot \sum n_i}.$$

By using these formulas, we derive an equation to calculate manganese molar fraction in the metal melt for the state of equilibrium:

$$\chi_{Mn}^{eq} = \frac{\% MnO_{eq} \cdot M_{FeO}}{K_{Mn} \cdot \% FeO_{eq} \cdot 2.5 \cdot 1.3 \cdot M_{MnO}}. \tag{5}$$

By assuming initial manganese and iron oxide concentrations close to equilibrium (*% Mn* = 25.3 and *% FeO* = 1.6), we obtain manganese molar fraction in the state of equilibrium, i.e. $\chi^{eq}_{Mn}$ ~ 0.59.

This method, however, has an obvious problem, which is the need to know activity factors ($\gamma_i$). This being said, it cannot be recommended for a reliable evaluation of equilibrium concentrations mainly because initial oxide concentrations can change considerably while reaching the equilibrium, especially if they are low (like the FeO concentration in our case). Moreover, this approach fails to take into account the masses of interacting phases.

Now let us consider a more complicated case of practical importance. We will attempt to calculate equilibrium concentrations for all reactants in reaction (1) allowing for the correlation between initial masses of the metal and oxide phases.

The following assumptions will be made:
Temperature: 1700°C;
Metal phase composition (mas. %) – 99 Fe, 1 Mn;
Oxide phase composition (mas. %): CaO – 3, MgO – 2, $Al_2O_3$ – 12, $SiO_2$ – 40, MnO – 40, FeO – 3;

Metal and oxide melt mass correlation: $\dfrac{m_{ox}}{m_{Me}} = 0.5$.

Chemical reaction in the metal/oxides boundary:

$$[Mn] + (FeO) \rightarrow (MnO) + [Fe]. \qquad (1)$$

Equilibrium constant:

$$K_p = \left( \dfrac{a_{MnO} \cdot a_{Fe}}{a_{Mn} \cdot a_{Fe}} \right)_{eq}$$

$a_{Fe} = 1,$

$$K_{Mn} = \dfrac{(MnO)_{eq}}{(FeO)_{eq} \cdot [Mn]_{eq}}. \qquad (6)$$

where parentheses - % mass of component in oxide phase,
square brackets - % mass of component in metal phase.
For example: T = 1700ºC, $K_{Mn} \sim 8.3$.

There are three unknowns: $(MnO)_{eq}$, $(FeO)_{eq}$, $[Mn]_{eq}$, therefore, we need three independent equations.

The first one will be an equilibrium constant equation (6), with the remaining two equations derived allowing for the balance of reactant masses. If manganese molar volume in the metal melt is 1 mole less, than MnO and FeO molar volume in the oxide melt will be 1 mole higher, as follows from stoichiometry of reaction (1). As an illustration, we obtain the following Mn molar volume variations as its concentration changes from initial *[Mn]₀* to equilibrium *[MnO]^{eq}*:

*[Mn]₀-[Mn]_{eq}*, in 100 g metal solution, g;

$\dfrac{[Mn]_0 - [Mn]_{eq}}{100}$ in 1 g of metal solution, g;

$\dfrac{([Mn]_0 - [Mn]_{eq}) \cdot m_{Me}}{100}$ in metal solution, g;

$$\frac{([Mn]_0 - [Mn]_{eq}) \cdot m_{Me}}{100 \cdot M_{Mn}} \text{ in metal solution, mol.}$$

Hence, balance equations will be as follows:

$$\frac{([Mn]_0 - [Mn]_{eq}) \cdot m_{Me}}{100 \cdot M_{Mn}} = \frac{((Mn)_{eq} - (Mn)_0) \cdot m_{Sl}}{100 \cdot M_{MnO}}; \qquad (7)$$

$$\frac{([Mn]_0 - [Mn]_{eq}) \cdot m_{Me}}{100 \cdot M_{Mn}} = \frac{((FeO)_0 - (FeO)_{FeO}) \cdot m_{Sl}}{100 \cdot M_{FeO}}. \qquad (8)$$

From the equations (6), (7) and (8) we can find equilibrium concentrations of Mn, FeO and MnO: $[Mn]_{eq} \sim 2$ mas.%, $(MnO)_{eq} \sim 35.6$ mas.%, $(FeO)_{eq} \sim 4.7$ mas.

As seen from the results, we have obtained not only the manganese equilibrium concentration, but also concentrations of $(FeO)_{eq}$ and $(MnO)_{eq}$ in the state of equilibrium, as opposed to the previous approach. It should be noted that the manganese equilibrium concentration is twice as high as the initial one.

Calculating equilibrium concentrations in the interaction among multicomponent phases is much more complicated, as the effect of parallel reactions needs tobe taken into account. Indeed, if a system is in the state of equilibrium, this state is reached by all potential reactions in the interphase boundary. The problem therefore is to calculate equilibrium concentrations for all and any components of the interacting phases.

**Calculating equilibrium composition of oxide and metal melts in a multicomponent system**

Multiple chemical reactions proceed in the interphase boundary (Fig.1). Hence, we need to know the equilibrium constant values for all potential processes, and also the details of material balance for all of the reactants. It should be noted that the choice of reactions is quite arbitrary. For example, the following reactions may proceed in the interphase boundary:

$$[Mn] + (FeO) \rightarrow [Fe] + (MnO), \qquad (9)$$

$$[Si] + 2(FeO) \rightarrow 2[Fe] + (SiO_2), \qquad (10)$$
$$[Si] + 2(MnO) \rightarrow 2[Mn] + (SiO_2) \qquad (11)$$

etc.

At the same time, independent reactions alone are to be taken into account here. In our case, for example, reaction (11) depends on reactions (9) and (10), all of its parameters being obtained by combining the parameters of (9) and (10). One of these three reactions therefore shall be disregarded in the analysis.

Now we will attempt to describe system equilibrium (Fig. 1) by assuming that one common reactant is involved in all of the reactions. For iron-based melts, this should be iron oxide (FeO), which is always present in oxide melts.

Chemical reactions of all common reactant-containing metal phase components in terms of 1 mole (FeO) are as follows:

$$[Mn] + (FeO) \rightarrow (MnO) + [Fe], \qquad (12)$$
$$½ [Si] + (FeO) \rightarrow ½(SiO_2) + [Fe], \qquad (13)$$
$$2/3 [Al] + (FeO) \rightarrow 1/3 \, Al_2O_3 + [Fe]. \qquad (14)$$

Or in general:
$$n/m \, [E_i] + (FeO) \rightarrow 1/m \, E_{in}O_m \qquad (15)$$

$a_{Fe} = 1$.

Therefore, there are $(2i + 1)$ unknowns:

$i$ equilibrium concentrations of reactants in the metal phase $(E_i)_{eq}$
$i$ equilibrium concentrations of reactants in the oxide phase $(E_{in}O_m)_{eq}$
and $1$ equilibrium concentration of common reactant $(FeO)$.

For the solution using the following $(2i + 1)$ equations:
$i$ equations for equilibrium constant:

$$K_{Ei} = \frac{(E_{in}O_m)_{eq}^{\frac{1}{m}}}{(FeO)_{eq}[E_i]_{eq}^{\frac{n}{m}}} \qquad (16)$$

$i$ equations of material balance, as shown above in (6), (7):

$$\frac{([E_i]_0 - [E_i]_{eq}) \cdot m_{Me}}{100 \cdot M_{Ei} \cdot n} = \frac{((E_{in}O_m)_{eq} - (E_{in}O_m)_o) \cdot m_{Sl}}{100 \cdot M_{E_{in}O_m}}. \qquad (17)$$

and *1* simple equation:

$$\Sigma(E_{in}O_m)_{eq} + (FeO)_{eq} + \Sigma(E_{in}X_m) = 100, \quad (18)$$

where $E_{in}X_m$ are oxides and other chemicals (sulphides, fluorides, etc.), which are not involved in chemical reactions.

This bulky set of equations shall be obviously solved by means of computer. One of the possible algorithms is shown below:

$$(E_{in}O_m)_{eq} = K_{E_i}^m \cdot (FeO)_{eq}^m \cdot [E_i]_{eq}^n; \quad (19)$$

$$\Sigma(K_{Ei}^m \cdot (FeO)_{eq}^m \cdot [E_i]_{eq}^n) + (FeO)_{eq} + \Sigma(E_{in}X_m)_{eq} = 100, \quad (20)$$

$$[E_i]_0 - [E_i]_{eq} = \frac{m_{sl}}{m_{Me}} \cdot \frac{n \cdot M_{E_i}}{M_{E_{i_n}O_m}} \cdot [K_{E_i}^m \cdot (FeO)_{eq}^m \cdot [E_i]_{eq}^n - (E_{i_n}O_m)_o]; \quad (21)$$

$$[E_i]_{eq} = \frac{\dfrac{m_{sl}}{m_{Me}} \cdot \dfrac{M_{E_i}}{M_{E_{i_n}O_m}} (E_{i_n}O_m)_0 + [E_i]_0}{1 + \dfrac{m_{sl}}{m_{Me}} \cdot \dfrac{M_{E_i}}{M_{E_{i_n}O_m}}} \cdot K_{E_i}^m (FeO)_{eq}^m. \quad (22)$$

Equilibrium concentrations of all metal melt components are calculated according to (22). By substituting (22) calculation results to (20), we can determine equilibrium concentrations of iron oxide $(FeO)_{eq}$, following which equilibrium concentrations of all oxide melt components $(E_{i\,n}O_m)_{eq}$ are calculated by using (19).

**Conclusions**

1. Calculation of metal and oxide melt equilibrium concentrations for specific reactions fails to accurately reflect a complex set of reactions among multicomponent phases.

2. A method to calculate phase equilibrium composition in a multicomponent system is based on applying one common reactant for all reactions involved.

3. Given that equilibrium is the only state possible at a constant temperature, taking into account all metal phase components would be sufficient to ensure correct calculations.

4. Benefits of the proposed method:

- choice of reactions can be arbitrary, providing that one common reactant is available;
- simulation process is quite simple;
- the number of components in the interacting phases is unlimited;
- this method enables evaluation of the threshold (equilibrium) phase condition in a complex technological process, and also determination of the vector of concentration changes of all of the components.

9. Sh. Banya, M. Hino, T. Nagasaka: 'Estimation of hydroxyl capacity of molten silicates by quadratic formalism based on the regular solution model'. Tetsu to Hagane 1993 79 (1) 26-33.
10. A. Pelton, P. Talley, R. Shama: 'Thermodynamic evaluation of phase equilibria in the $CaCl_2$-$MgCl_2$-$CaF_2$-$MgF_2$ system'. J. Phase Equilib. 1992 13 (4) 384-390.
11. T. Matsumia, A. Nogami, H. Savada: 'Monte Carlo simulations of intermetallic compounds'. Adv. Mater. Processes. 1995 147 (2) 51-53.
12. T. Iita, Y. Kita, H. Okano, I. Katayama, T. Tanaka: 'An equation for the vapor pressure of liquid metals and calculation of their enthalpies of evaporation'. High Temp. Mater. Processes. 1992 10 (4) 199-207.
13. Y. Zuo, Y. Chang: 'Calculation of phase diagram and solidification paths of aluminium-rich aluminium- magnesium-copper ternary alloys'. Light Metals 1993 935-942.
14. P. Wu, G. Eriksson, A. Pelton, M. Blander: 'Prediction of the thermodynamic properties and phase diagrams of silicate systems - Evaluation of the FeO-Mg-$SiO_2$ system'. ISIJ Int. 1993 33 (1) 26-35.
15. M. Blander, I. Bloom: 'A statistical mechanical theory for molten silicate solutions'. Proc. Electrochemic. Soc. 1994 -13 1-7.
16. W. Van Niekerk, R. Dippenaar: 'Thermodynamic aspects of $Na_2O$ and $CaF_2$ containing lime-based slags used for the desulphurization of hot-metal'. ISIJ Int. 1993 33 (1) 59-65.
17. Y. Xiao, L. Holappa: 'Determination of activities in slags containing chromium oxides'. ISIJ Int. 1993 33 (1) 66-74.
18. J.M. Dowden, R. Ducharme, P. Kapadia and A. Clucas: 'A mathematical model for the penetration depth in welding with continuous $CO_2$ lasers.' In Proc. ICALEO'94 Laser Inst. America 1995 79 451-460.
19. V. Selivorstov, Y. Dotsenko, K. Borodianskiy: 'Gas-dynamic influence on the structure of cast of A356 alloy'. Herald of the Donbass State Engineering Academy. Collection of science papers, 3 (20) 2010 234-238.
20. M. Zinigrad, V. Mazurovsky, K. Borodianskiy: 'Physico-Chemical and Mathematical Modeling of Phase Interaction taking place
270